\def\beq#1{\begin{equation}\label{#1}}
\def\eeq{\end{equation}}
\def\beqa#1{\begin{eqnarray}\label{#1}}
\def\eeqa{\end{eqnarray}}

\def\fun#1#2{\lower3.6pt\vbox{\baselineskip0pt\lineskip.9pt
        \ialign{$\mathsurround=0pt#1\hfill##\hfil$\crcr#2\crcr\sim\crcr}}}
        
\def\xi{{{\bf x}^b}}

\documentclass[twocolumn,aps,showpacs,showkeys,nofootinbib]{revtex4}
\usepackage{epsfig}

\newcommand{\be}{\begin{equation}}
\newcommand{\ee}{\end{equation}}
\newcommand{\ba}{\begin{eqnarray}}
\newcommand{\ea}{\end{eqnarray}}

\begin{document}
\input{epsf.sty}

\title{A parametric reconstruction of the cosmological jerk from diverse observational data sets}
\author{Ankan Mukherjee$^{1}$ \footnote{E-mail: ankan\_ju@iiserkol.ac.in} and Narayan Banerjee$^{2}$ \footnote{E-mail: narayan@iiserkol.ac.in}}
\address{$^1$$^{,2}$~Department of Physical Sciences,~~\\Indian Institute of Science Education and Research Kolkata,~~\\ Mohanpur, West Bengal-741246, India.}

\begin{abstract}
A parametric reconstruction of the jerk parameter, the third order derivative of the scale factor expressed in a dimensionless way, has been discussed. Observational constraints on the model parameters have been obtained by Maximum Likelihood Analysis of the models using Supernova Distance Modulus data (SNe), Observational Hubble Data (OHD), Baryon Acoustic Oscillation (BAO) data and CMB shift parameter data (CMBShift). The present value of the jerk parameter has been kept open to start with, but the plots of various cosmological parameter like deceleration parameter $q(z)$, jerk parameter $j(z)$, dark energy equation of state parameter $w_{DE}(z)$ indicate that the reconstructed models are very close to a $\Lambda$CDM model with a slight  inclination towards a non-phantom behaviour of the evolution. 
\end{abstract}



\maketitle


\section{Introduction}

Hubble's discovery of the expansion of the universe certainly is the foundation of the development of modern cosmology as an observational science. As this discovery shows that the universe indeed has an expansion, cosmologists naturally were interested in the evolution of Hubble's parameter $H$ which is the fractional rate of the expansion of the universe. Evolution of $H$ is indicated by a deceleration parameter $q$, believed to be constant until recently. Now that one is convinced about the evolution of the deceleration parameter through the observational evidence of the universe entering into an accelerated phase of expansion from a decelerated one in the recent past, the subject of interest is expected to be to look at the evolution of $q$, i.e., some third derivative of the size of the universe. \\

After the discovery of the accelerated expansion of the universe\cite{Riess,Perlmutter,schmidt,Knop,tonry,barris,hicken,suzuki}, and the absence of any observational detection of any matter responsible for this unexpected behaviour, various models for such a matter has been proposed. There are comprehensive reviews\cite{varun1,varun2,sean,ratra,paddy,sami} that deal with the merits of various options and also the challenges they have to encounter. Modifications of the theory of gravity so as to give room for the accelerated expansion have also been attempted. An inclusion of nonlinear terms of the Ricci scalar in the action\cite{Capozziello,Carroll,sudipta, mena, nojiri}, higher dimensional theories of gravity\cite{deffayet,dvali}, nonminimally coupled scalar field theory like Brans-Dicke theory\cite{nb} are amongst some of them. \\

As there is no theoretical compulsion of any of the models from other branches of physics, such as particle physics theory, there are attempts to find some viable models right from the observations. The basic idea is to write down the evolution pattern  in the Einstein's field equations for a spatially homogeneous and isotropic model and to find out the field, such as the scalar field potential\cite{em}, giving rise to that kind of an evolution. This kind of a reverse engineering, called a ``reconstruction'', finds a very wide application in recent literature in cosmology. Starobinsky showed that the potential for a scalar field can be determined by using the data on density perturbation\cite{staro}. Huterer and  Turner\cite{huterer1, huterer2} utilzed the data of distance measurement for the purpose of such a reconstruction. \\

Reconstruction of a dark energy (DE), that drives the current acceleration of the universe, normally involves finding the equation of state parameter $w_{DE}$ of the DE given by $ w_{DE}=\frac{p_{\phi}}{\rho_{\phi}}$, where $p_{\phi}$ and ${\rho}_{\phi}$ represent the contribution to the  pressure and the density respectively by the DE\cite{varun2, saini1}. This is done in two different ways, one is called a parametric reconstruction where a suitable ansatz for $w_{DE} =w_{DE}(z)$ (where $z$ is the redshift given by $1+z = \frac{a_{0}}{a}$) is chosen and the values of the parameters in the anstaz are estimated with the help of obseravational data\cite{gerke,gong,holsclaw, polarski, linder, anjan1, anjan2}. The other one, a non-parametric reconstruction, is an attempt to build up the actual functional form of $w_{DE}$ directly from the available data\cite{hu, slp1, slp2, czp}. Pan and Alam discussed the use of different cosmological parameters in testing the viability of different reconstructed dark energy models\cite{pa}.\\

In building up the model by the reverse engineering, i.e., a reconstruction, attempts are normally through the dark energy equation of state. There are only a few attempts to find a suitable model by a reconstruction of the kinematical quantities like the deceleration or higher order derivatives of the scale factor. Gong and Wang made an attempt to exercise a reconstruction of  the deceleration parameter $q$\cite{gong}. Another such attempt is by Wang {\it et al}\cite{wang}  where the dark energy potential is obtained via a reconstruction of deceleration parameter. \\

In spite of its being the natural choice amongst the kinematical quantities, as discussed at the beginning, the jerk parameter $j$, given by $j =-\frac{1}{a H^{3}}\frac{d^{3}a}{dt^{3}}$ had hardly been investigated in the context of the accelerated expansion. Visser\cite{matt} initiated a serious discussion on the jerk parameter, albeit from a different motivation. The idea was to look at the equation of state of the cosmic fluid as a Taylor expansion in terms of a background density. Analytical expressions for jerk and parameters constructed by involving higher order derivatives of the scale factor in various forms of matter in a Friedmann universe had been presented by Dunajski and Gibbons\cite{gary}. \\

A reconstruction of the dark energy equation of state through a parametrization of the jerk parameter has been very recently given by Luongo\cite{luongo}.  A systematic study of jerk as the way towards building up a model for the accelerated expansion was given by Rapetti {\it et al}\cite{rapetti}. In an  exhaustive recent work by Zhai {\it et al}\cite{zz}, a reconstruction of $j$ has been attempted. For four different forms of $j = j(z)$, they fixed the parameters in $j$ by using Observational Hubble parameter data (OHD) and Union 2.1 Supernovae (SNe) data\cite{suzuki}. \\

It should be clearly mentioned that the equation of state parameter $w_{DE}$, the energy density of the dark energy, the potential of the quintessence field etc. are all part of the theoretical input, and hence constitute the fundamental ingredients of the model. The deceleration parameter, the jerk etc. are kinematical quantities, and thus are the outcome of the solution of the system of equations of the model. So no wonder that a reconstruction of $w_{DE}$ holds the centre-stage in attempts towards building up a model for the accelerating universe. It should be mentioned that the basic advantage of this kind of reconstruction through kinematical quantities neither assumes any theory of gravity (like GR, f(R) gravity, scalar-tensor theory etc.) nor does it assume a given matter distribution like a quintessence field or any other exotic matter though any equation of state. Thus, this reconstruction may lead to some understanding of the basic matter distribution and the possible interaction amongst them without any assumptions on them to start with.\\

The motivation of the present work is to reconstruct a dark energy model through the jerk parameter $j$. If $j$ is known as $j(t)$ or $j(z)$, the third order differential equation for the scale factor is known and hence one can find the evolution, at least in principle. It is wellknown that the $\Lambda$CDM model does very well in explaining the present acceleration of the universe but fails to match the theoretically predicted value of $\Lambda$. Thus the attempts to build up dark energy models hover around finding one which in the limit of $z=0$ (the present epoch) resembles a $\Lambda$CDM model. The work by Zhai {\it et al}\cite{zz} is also along that line. For a $\Lambda$CDM model, one has a constant $j$ as $j=-1$. Zhai  {\it et al} wrote their ansatz in such a way that the present value of $j$ is actually $-1$, i.e., $j(z=0) = -1$. All the four different forms of $j=j(z)$, they wrote, have this feature. \\

The present work is more general in two different ways. First, the functional form of $j=j(z)$ in this work is more relaxed in the sense that its present value depends on a parameter to be fixed by observations, and is allowed to have a value widely different from $-1$ if so required by the data sets. In this work, different ansatzs for $j=j(z)$ are taken, all of which are quite free to take values very much different from $-1$. The second point is related to sophistication in the sense that where Zhai {\it et al} used a combination of two data sets namely the OHD and the SNe Ia, we have a combination of four data sets. In addition to the two used by  Zhai {\it et al}, the Baryon Acoustic Oscillation (BAO) data and the very recent CMB shift parameter (CMBShift) data have also been incorporated in the present work. As a result, we are able to restrict or constrain the parameters of the theory to narrower limits. The results obtained by the present work also show that the observations very strongly indicate an inclination towards very close to a $\Lambda$CDM behaviour of the present distribution of matter with a marginal preference towards a non-phantom behaviour. \\

The paper is arranged as the following. Section 2 presents the basic formalism of FRW cosmology along with the definitions of different kinematical quantities. In section 3, different observational data, used in the present work, have been briefly discussed. Reconstruction of four different models for jerk parameter have been presented in section 4 along with the statistical analysis. In the fifth section, Bayesian analysis regarding the model selection has been discussed. The sixth and final section contains a discussion on the results obtained.

\section{Kinematical quantities}
The metric for a homogeneous and isotropic universe with a flat spatial section is written as 
\be
ds^2=-dt^2+a^2(t)(dr^2+r^2d\Omega^2),
\label{FRWmetric}
\ee 
where  $a(t)$, the time dependent coefficient of the space part, is called the scale factor.  It  takes care of the time dependence of the spatial separation between two objects (galaxies, in the context of the present universe) in the cosmological scale. The kinematical quantities are usually defined as the time derivatives of $a$ in a dimensionless way. It is convenient to use the redshift $z$  as the argument instead  of cosmic time $t$  as $z$ is dimensionless and is an observational quantity.  Redshift $z$ is defined as $(1+z)=\frac{a_0}{a(t)}$, where $a_0$ is the  present value of the scale factor. The Hubble parameter ($H$) is defined as $H(t)=\frac{\dot{a}}{a}$, where a dot indicates derivative with respect to $t$ and this $H(t)$ can also be written as a function of the redshift $z$, i.e. $H=H(z)$. The other kinematical quantities, related to the expansion of the universe, are the higher order derivatives. The deceleration parameter $q(z)$ and the jerk parameter $j(z)
 $ can be written in terms of $H(z)$ with $z$ as the argument as

\be
q(z)=-\frac{1}{H^2a}\frac{d^2a}{dt^2}=-1+\frac{1}{2}(1+z)\frac{[H^2(z)]'}{H^2(z)},
\label{decelerationparameter}
\ee
\be
j(z)=-\frac{1}{H^3a}\frac{d^3a}{dt^3}=-1+(1+z)\frac{[H^2(z)]'}{H^2(z)}-\frac{1}{2}(1+z)^2\frac{[H^2(z)]''}{H^2(z)},
\label{jerkparameter}
\ee
where a prime denotes the derivative with respect  to the  redshift $z$. One can extend this chain of derivatives, such as the fourth order derivative, called the snap parameter $s$ can be defined as $s = \frac{1}{a H^{4}}\frac{d^{4} a}{d t^{4}}$ and so on \cite{matt}. We shall however, restrict to the third derivative, namely the jerk parameter, as the evolution of $q$ is of physical interest now. It is to be noted that in defining $j$, there are two conventions of using a positive or a negative sign. We have adopted the convention as used by Zhai {\it et al}\cite{zz} and not the one used by Visser\cite{matt} and Luongo\cite{luongo}. This preference is only because we shall compare our results with that of Zhai {\it et al}\cite{zz}.

\section{Observational data} 
 Here we have used some recent data sets for the reconstruction, namely, \\
\begin{enumerate}
\item{Observational Hubble parameter data (OHD),} 
\item{Type Ia Supernova data (SNe),}
\item {Baryon Acoustic Oscillation  data (BAO),}
\item {CMB shift parameter data (CMBShift)}. 
\end{enumerate}
In the following, these four datasets are very briefly described.

\subsection{Observational Hubble parameter data (OHD):}
We have used the observational data of Hubble parameter $H(z)$ measured  by different groups. The estimation of the value of $H(z)$ can be obtained from the measurement of differential of redshift $z$ with  respect to cosmic time $t$ as
\be
H(z)=-\frac{1}{(1+z)}\frac{dz}{dt}.
\label{Hz}
\ee
Simon {\it et al.} have used differential age of galaxies as an estimator of $dz/dt$ \cite{simon} and this data has been utilized in the present work. Measurement of cosmic expansion history using red-enveloped galaxies was done by Stern {\it et al} \cite{stern}. In addition to that, compilation of observational Hubble parameter measurement has been presented by Moresco {\it et al} \cite{moresco}. The differential age method along with SDSS data have been adopted by Zhang {\it et al}  so as to measure Hubble parameter at low redshift \cite{zhang}. The  model parameter values can be estimated using $\chi^2$-statistics, defined as
\be
\chi^2_{{\tiny OHD}}=\sum_{i}\frac{[H_{obs}(z_i)-H_{th}(z_i,\{\theta\})]^2}{\sigma_i^2},
\label{chiOHD}
\ee 
where $H_{obs}$ is the observed value of the Hubble parameter, $H_{th}$ is theoretical one and $\sigma_i$ is the uncertainty  associated to the $i^{th}$ measurement. The $\chi^2$ is a function of the set of model parameters $\{\theta\}$.

\subsection{Type Ia Supernova data:}
Type Ia Supernova data is actually the difference between the apparent magnitude ($m_B$) and absolute magnitude ($M_B$) in the B-band of the observed type Ia supernova, named as distance modulus ($\mu_B$), and is defined as 
\be 
\mu_B(z)=5\log_{10}\Bigg(\frac{d_L(z)}{1Mpc}\Bigg)+25,
\label{distmodulus}
\ee
where $d_L(z)$ is the luminosity distance. In a spatially flat FRW universe, luminosity distance $d_L(z)$ is defined as
\be
d_L(z)=(1+z)\int_0^z\frac{dz'}{H(z')}.
\label{lumdist}
\ee  
Here the 31 binned distance modulus data sample of joint lightcurve analysis (jla) \cite{jlaBetoule} has been used. To introduce the correlation matrix of the binned data sample, the method discussed by Farooq, Mania and Ratra \cite{faroqmaniaratra} has been adopted. The $\chi_{SNe}^2$ has been defined as
\be
\chi_{SNe}^2=A(\{\theta\})-\frac{B^2(\{\theta\})}{C}-\frac{2\ln{10}}{5C}B(\{\theta\})-Q,
\ee     
where 
\be 
A(\{\theta\})=\sum_{\alpha,\beta}(\mu_{th}-\mu_{obs})_{\alpha}(Cov)^{-1}_{\alpha\beta}(\mu_{th}-\mu_{obs})_{\beta},
\ee
\be
B(\{\theta\})=\sum_{\alpha}(\mu_{th}-\mu_{obs})_{\alpha}\sum_{\beta}(Cov)^{-1}_{\alpha\beta},
\ee
\be
C=\sum_{\alpha,\beta}(Cov)^{-1}_{\alpha\beta},
\ee
and the $Cov$ is the $31\times31$ covarience matrix of the binned data. Here the $Q$ is a constant which does not depend upon the parameters and hence has been ignored. The technical details of binning the data and handling the other parameters involved, like the stretch, colour and reference magnitude has been discussed comprehensively in reference \cite{jlaBetoule}.

\subsection{Baryon Acoustic Oscillation (BAO) data:}
We also utilize the Baryon acoustic oscillation (BAO) data along with the ``acoustic scale ($l_A$)", the comoving sound horizon ($r_s$) at photon decoupling epoch ($z_*$) and at drag epoch ($z_d$) as measured by Planck\cite{planck, wangwang}. The comoving sound horizon at photon decoupling is defined as
\be
r_s(z_*)=\frac{c}{\sqrt{3}}\int_0^{1/(1+z_*)}\frac{da}{a^2H(a)\sqrt{1+a(3\Omega_{b0}/4\Omega_{\gamma 0})}},
\label{rs}
\ee
where $\Omega_{b0}$ is the present value of Baryon density parameter and $\Omega_{\gamma 0}$ is the present value of photon density parameter. According to the Planck results, the value of redshift at photon decoupling is $z_*\approx 1091$ and reshift at drag epoch is $z_d\approx 1021$ \cite{planck}.  The acoustic scale at decoupling is defined as
\be
l_A=\pi\frac{d_A(z_*)}{r_s(z_*)}
\label{accscale}
\ee
where $d_A(z_*)=c\int_0^{z_*} \frac{dz'}{H(z')}$, is the comoving angular diameter distance at decoupling. Another important definition  is that of ``dilation scale", given by $D_V(z)=[czd_A^2(z)/H(z)]^{\frac{1}{3}}$. Here we have taken three mutually uncorrelated measurements of $\frac{r_s(z_d)}{D_V(z)}$, the result of 6dF galax survey at redshift $z=0.106$ \cite{beutler}, and the results of Baryon Oscillation Spectroscopic Survey at redshift $z=0.32$ (BOSS LOWZ) and at redshift $z=0.57$ (BOSS CMASS) \cite{landerson}. Combining these data with Planck results \cite{planck, wangwang}, given as $l_A=301.74\pm0.19$, $\frac{r_s(z_d)}{r_s(z_*)}=1.019\pm0.009$, we finally obtained the ratio $\Big(\frac{d_A(z_*)}{D_V(z_{BAO})}\Big)$ at three different values of $z_{BAO}$. These can be utilized as the BAO/CMB constraint for dark energy models. Table \ref{tableBAO} contains the values of $\Big(\frac{r_s(z_d)}{D_V(z_{BAO})}\Big)$ and that of $\Big(\frac{d_A(z_*)}{D_V(z_{BAO})}\Big)$ as
discussed.

\begin{table}[h!]
\begin{center}
\resizebox{0.49\textwidth}{!}{  
\begin{tabular}{ |c ||c |c |c |} 
 \hline
 $z_{BAO}$ & 0.106 & 0.32 & 0.57 \\ 
 \hline
 \hline
 $\frac{r_s(z_d)}{D_V(z_{BAO})}$ & 0.3228$\pm$0.0205 & 0.1167$\pm$0.0028 & 0.0718$\pm$0.0010 \\ 
 \hline
 $\frac{d_A(z_*)}{D_V(z_{BAO})}\frac{r_s(z_d)}{r_s(z_{*})}$ & 31.01$\pm$1.99 & 11.21$\pm$0.28  & 6.90$\pm$0.10 \\ 
 \hline
 $\frac{d_A(z_*)}{D_V(z_{BAO})}$ & 30.43$\pm$2.22 & 11.00$\pm$0.37  & 6.77$\pm$0.16\\ 
 \hline
\end{tabular}
}
\end{center}
\caption{BAO/CMB data table \cite{planck,wangwang,beutler,landerson}}.
\label{tableBAO}
\end{table}

The relevant $\chi^2$, namely $\chi^2_{BAO}$, is defined as:
\be
\chi^2_{BAO}={\bf X^{t}C^{-1}X},
\label{chibao}
\ee
where
\[
{\bf X}=
\left( {\begin{array}{c}
   \frac{d_A(z_*)}{D_V(0.106)}-30.43   \\
   \frac{d_A(z_*)}{D_V(0.2)}-11.00     \\
   \frac{d_A(z_*)}{D_V(0.35)}-6.77  \\
\end{array} } \right)
\]
and $C^{-1}$, the inverse of the covariance matrix. As the three measurements are mutually uncorrelated, the covariance matrix is diagonal. \\

We had to resort to a mixing of the BAO data with the CMB measurements in order to remove the dependence on sound horizon. There are of course other ways of doing that, which might lead to some other issues. For instance, if the acoustic scale is used separately, one would need to use the present baryon density parameter (${\Omega}_{b0}$) and the photon density paramater (${\Omega}_{\gamma 0}$) which are model dependent to a large extent.   \\
A detailed discussion regarding the statistical analysis of cosmological models using BAO data is available in reference  \cite{giostri}.  

\subsection{CMB shift parameter (CMBShift) data:}
The CMB shift parameter is related to the position of the first acoustic peak in power spectrum of the temperature anisotropy of the Cosmic Microwave Background (CMB) radiation. The value of CMB shift parameter is not directly measured from CMB observation. The value is estimated from the CMB data along with some assumption about the model of the background cosmology. For a spatially flat universe, the CMB shift parameter is defined as:
\be
{\mathcal R}=\sqrt{\Omega_{m0}}\int_0^{z_*}\frac{dz}{h(z)},
\ee 
where $\Omega_{m0}$ is the matter density parameter, $z_*$ is the redshift at photon decoupling and $h(z)=\frac{H(z)}{H_0}$ (where $H_0$ is the present value of Hubble parameter). The $\chi^2_{\tiny CMBShift}$ is defined as 
\be 
\chi^2_{\tiny CMBShift}=\frac{({\mathcal R}_{obs}-{\mathcal R}_{th}(z_*,\{\theta\}))^2}{\sigma^2},
\ee
where ${\mathcal R}_{obs}$ is the observationally estimated value of CMB shift parameter and $\sigma$ is the corresponding uncertainty. In this work, the value of CMB shift parameter as estimated from Planck data \cite{wangwang} has been used.

\section{Reconstruction of Jerk parameter}
The evolution of a spatially flat homogeneous and isotropic universe is governed by Einstein's field equations, given by 
\be 
3H^2=8\pi G(\rho_m+\rho_{DE}),
\label{friedmann1}
\ee
\be
2\dot{H}+3H^2=-8\pi Gp_{DE},
\label{friedmann2}
\ee
where $\rho_m$ is the density of the pressureless dust matter, $\rho_{DE}$ and $p_{DE}$ are the contribution of the dark energy to the energy density and pressure sector. An overheaded dot represents the differentiation with respect to the time. \\ 
The jerk parameter $j(z)$, defined in equation \ref{jerkparameter}, is the dimensionless representation of the third order time derivative of the scale factor $a(t)$. For $\Lambda$CDM cosmology, jerk parameter is a constant with the value $j_{\Lambda CDM}=-1$. We follow the parametric reconstruction of $j(z)$ in the similar line as discussed by Zhai {\it et al} \cite{zz}. The major difference, as discussed before, is that we do not restrict $j=-1$ for $z=0$. So we do not assume a $\Lambda$CDM for the present universe {\it a priori}, but rather allow the model to behave in a more general way. The present value of $j$ is allowed to be fixed by the observational data.  We write the jerk parameter is as
\be
j(z)=-1+j_1\frac{f(z)}{h^2(z)},
\ee 
where $j_1$ is a constant, $h(z)=\frac{H(z)}{H_0}$ is the Hubble parameter scaled by its present value and $f(z)$ is an analytic function of $z$. Four ansatz for $j(z)$ have been chosen in the present work, which will be discussed separately. Here $j_1$ is the model parameter to be fixed by observational data. Model I is the one where the evolution of $j$ is taken care of solely by $h^{2}(z)$. For the other three, the redshift $z$ also contributes explicitly and not through $h^2(z)$ alone. The four ansatz chosen are given below,
\be
Model ~I.~~~~   j(z)=-1+j_1\frac{1}{h^2(z)},
\label{jmodel1}
\ee
\be
Model ~II.~~~~~  j(z)=-1+j_1\frac{(1+z)}{h^2(z)},
\label{jmodel2}
\ee
\be
Model ~III.~~~~~  j(z)=-1+j_1\frac{(1+z)^2}{h^2(z)},
\label{jmodel3}
\ee
\be
~~Model ~IV.~~~~~   j(z)=-1+j_1\frac{1}{(1+z)h^2(z)}.
\label{jmodel4}
\ee

By substituting these expressions in the definition of $j$ given by equation (\ref{jerkparameter}), one can get the solutions for $h^{2}$ as first integrals as 
\be
Model~I. ~~~~h^2(z)=\frac{H^2(z)}{H_0^2}=c_1(1+z)^3+c_2+\frac{2}{3}j_1ln(1+z),
\label{jmodelh1}
\ee  
\be
Model~II.~~~~h^2(z)=\frac{H^2(z)}{H_0^2}=c_1(1+z)^3+c_2+j_1(1+z),
\label{jmodelh2}
\ee
\be
Model~III.~~~~h^2(z)=\frac{H^2(z)}{H_0^2}=c_1(1+z)^3+c_2+j_1(1+z)^2,
\label{jmodelh3}
\ee
\be
Model~IV.~~~~h^2(z)=\frac{H^2(z)}{H_0^2}=c_1(1+z)^3+c_2-j_1\frac{1}{2(1+z)}.
\label{jmodelh4}
\ee
Here $c_1$ and $c_2$ are integration constants. Now from the scaling $h^2(z)=\frac{H^2(z)}{H_0^2}$, one has the constraint $h(z=0)=1$, which connects the constants as
\be
Model~I. ~~~~~~c_2=1-c_1,
\label{c2model1}
\ee  
\be
Model~II.~~~~~~~~c_2=1-j_1-c_1,
\label{c2model2}
\ee
\be
Model~III.~~~~~~~c_2=1-j_1-c_1,
\label{c2model3}
\ee
\be
Model~IV.~~~~~~~~c_2=1+\frac{j_1}{2}-c_1.
\label{c2model4}
\ee
Thus each of the models have two independent parameters, $(c_1,j_1)$. It is quite apparent from the expression of $h^2(z)$, through the Einstein's field equation, $3H^{2} = 8\pi G \rho$, that the parameter $c_1$ is equivalent to the matter density parameter $\Omega_{m0}$ which is the ratio of present matter density ($\rho_{m0}$) and critical density ($\rho_c=3H_0^2/8\pi G$).

\par The properties of dark energy can also be ascertained to a certain extent from the analytic expressions of the Hubble parameter for the models. As the dust matter is minimally coupled to the dark energy for these models, the dark energy density scaled by critical density can be expressed as
\be
\frac{\rho_{DE}}{\rho_c}=h^2(z)-c_1(1+z)^3.
\label{rhoDE}
\ee
The contribution of dark energy to the pressure sector can be obtained from equation (\ref{friedmann2}) as 
\be
\frac{p_{DE}}{\rho_c}=\frac{2}{3}(1+z)hh'-h^2(z).
\label{pDE}  
\ee
And finally the dark energy equation of state parameter for the models can be expressed as 
\be
w_{DE}(z)=\frac{\frac{2}{3}(1+z)hh'-h^2}{h^2(z)-c_1(1+z)^3}.
\label{wDE}
\ee
The analytic expressions of $w_{DE}$ obtained for the models are 
\be
Model~I. ~~~w_{DE}=-1+\frac{\frac{2}{9}j_1}{\frac{2}{3}j_1\ln{(1+z)}+(1-c_1)},
\label{wDEmodel1}
\ee  
\be
Model~II.~~~w_{DE}=-1+\frac{\frac{1}{3}j_1(1+z)}{j_1(1+z)+(1-c_1-j_1)},
\label{wDEmodel2}
\ee
\be
Model~III.~~~w_{DE}=-1+\frac{\frac{2}{3}j_1(1+z)^2}{j_1(1+z)^2+(1-c_1-j_1)},
\label{wDEmodel3}
\ee
\be
Model~IV.~~~w_{DE}=-1+\frac{\frac{1}{6}\frac{j_1}{(1+z)}}{(1-c_1+\frac{1}{2}j_1)-\frac{j_1}{2(1+z)}} .
\label{wDEmodel4}
\ee

\par The functional form of some of these expressions for $w_{DE}$ are indeed similar to some models existing in the literature. The values of the parameters are however quite different. For instance, Model II and IV indicate that $w_{DE}$ is a ratio of linear functions of $z$. Such an expression is quite widely used in the literature, for example, by Linder\cite{linder}, Gong and Wang\cite{gong}. The latter also contains a $w_{DE}$ which is a ratio of two quadratic functions of $z$ which is functionally similar the present model III.

\par The statistical analysis have been carried out using various combinations of the SNe, OHD, BAO and CMBShift data. The constraints on the parameters are obtained by the $\chi^2$ minimization. This is equivalent to the Maximum Likelihood analysis. This has been done numerically using the basic grid searching of likelihood where the range of parameters are divided into grids and all possible combinations are evaluated to estimate the maximum likelihood. The likelihood function, which is proportional  to the posterior probability distribution of the model parameters with a flat prior assumption, is defined as
\be
{\mathcal L}=\exp{\Big(-\frac{\chi^2}{2}\Big)}.
\ee
The combined $\chi^2$ is defined as 
\begin{equation}
\chi_{\tiny total}^2=\sum_i\chi^2_i,
\end{equation}
where $i$ indicates the data sets in the combination ($i=SNe, OHD, BAO, CMBShift$).

\par
\begin{figure}[htb]
\begin{center}
\includegraphics[angle=0, width=0.2\textwidth]{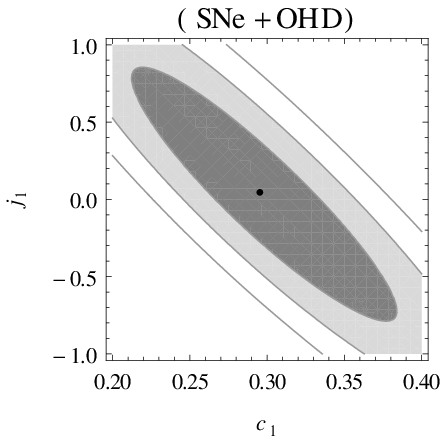}
\includegraphics[angle=0, width=0.2\textwidth]{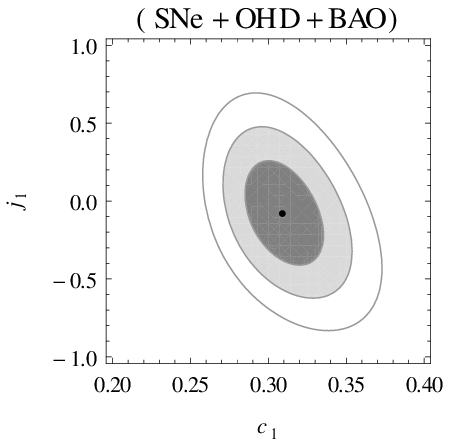}\\
\includegraphics[angle=0, width=0.2\textwidth]{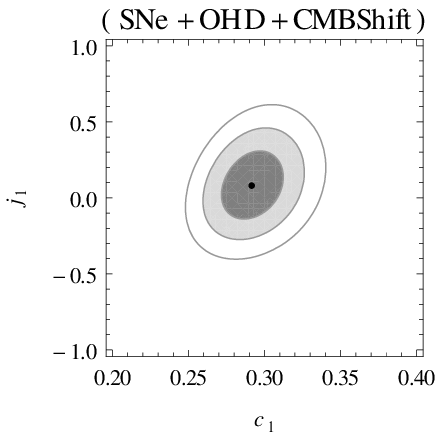}
\includegraphics[angle=0, width=0.2\textwidth]{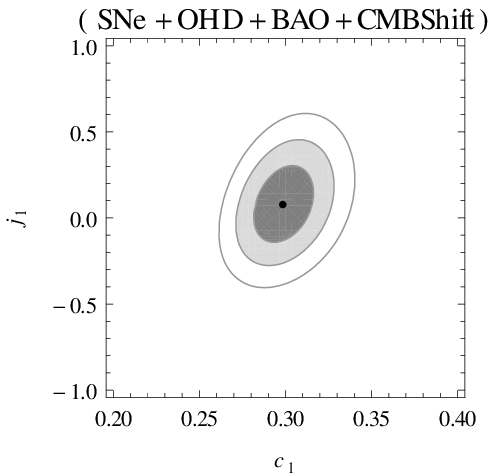}
\end{center}
\caption{{\small The confidence contours on 2D parameter space of model I. The 1$\sigma$, 2$\sigma$ and 3$\sigma$ confidence contours are presented from inner to outer regions and the central black dots represents the corresponding best fit points. The upper left panel is obtained for (SNe+OHD), upper right panel is for (SNe+OHD+BAO), lower left panel is for (SNe+OHB+CMBShift) and lower right panel is for (SNe+OHD+BAO+CMBShift).}}
\label{jz1contours}
\end{figure}

\begin{figure}[htb]
\begin{center}
\includegraphics[angle=0, width=0.2\textwidth]{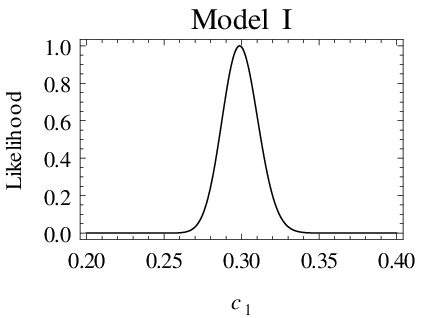}
\includegraphics[angle=0, width=0.2\textwidth]{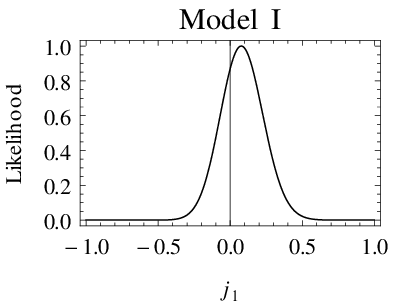}
\end{center}
\caption{{\small The marginalised likelihood functions of the model I obtained for SNe+OHD+BAO+CMBShift.}}
\label{jz1likelihood}
\end{figure}

 Figure \ref{jz1contours} shows the confidence contours on the 2D parameter space ($c_1$,$j_1$) of model I obtained for different combinations of the data sets. Figure \ref{jz1likelihood} presents the marginalised likelihood as functions of the model parameters $c_1$ and $j_1$ for model I obtained for SNe+OHD+BAO+CMBShift. The likelihood functions are well fitted to Gaussian distribution function with the best-fit parameter values $c_1=0.298\pm0.010$ and $j_1=0.078\pm0.140$. Table \ref{tablejzM1} presents the results of statistical analysis for model I. It is clear from the results that the addition of CMB shift parameter data leads to a substantial improvement of the parameter constraints.

\par
\begin{figure}[tb]
\begin{center}
\includegraphics[angle=0, width=0.2\textwidth]{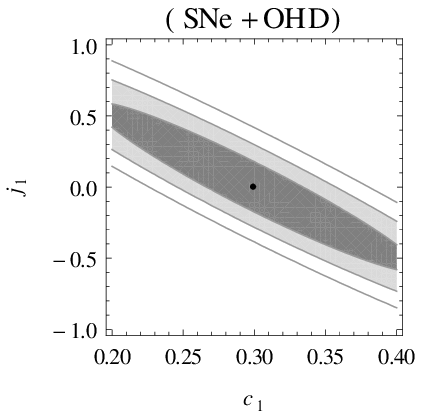}
\includegraphics[angle=0, width=0.2\textwidth]{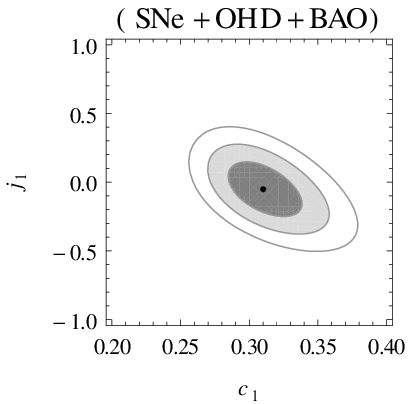}\\
\includegraphics[angle=0, width=0.2\textwidth]{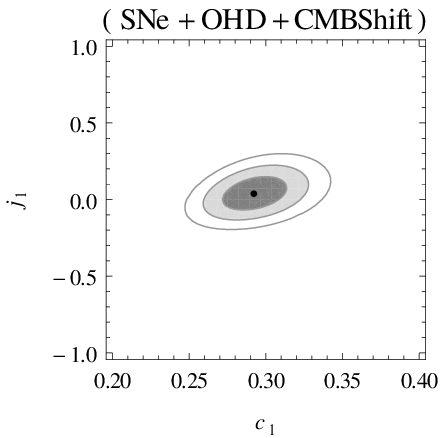}
\includegraphics[angle=0, width=0.2\textwidth]{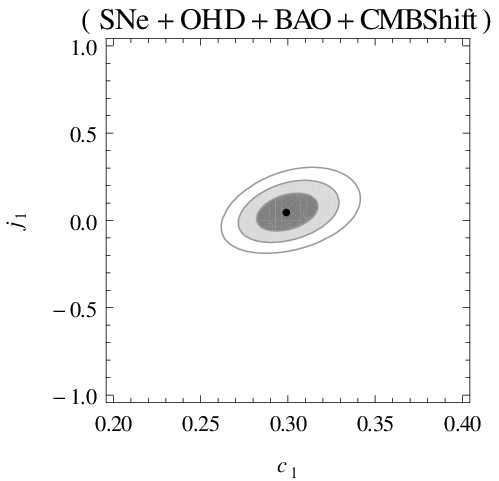}
\end{center}
\caption{{\small The confidence contours on 2D parameter space of model II. The 1$\sigma$, 2$\sigma$ and 3$\sigma$ confidence contours are presented from inner to outer regions and the central black dots represents the corresponding best fit points. The upper left panel is obtained for (SNe+OHD), upper right panel is for (SNe+OHD+BAO), lower left panel is for (SNe+OHB+CMBShift) and lower right panel is for (SNe+OHD+BAO+CMBShift).}}
\label{jz2contours}
\end{figure}

\begin{figure}[tb]
\begin{center}
\includegraphics[angle=0, width=0.2\textwidth]{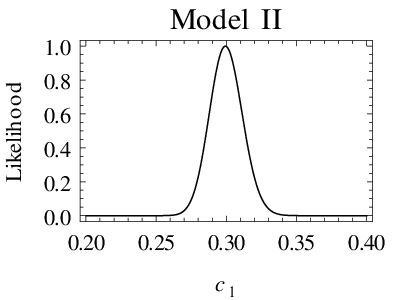}
\includegraphics[angle=0, width=0.2\textwidth]{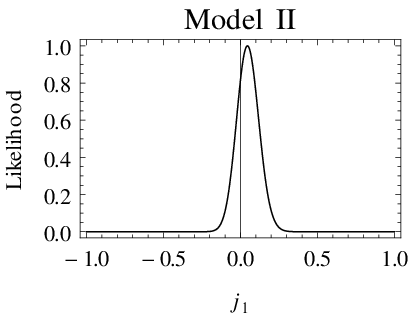}
\end{center}
\caption{{\small  The marginalised likelihood functions of the model II obtained for SNe+OHD+BAO+CMBShift.}}
\label{jz2likelihood}
\end{figure}

\begin{figure}[tb]
\begin{center}
\includegraphics[angle=0, width=0.2\textwidth]{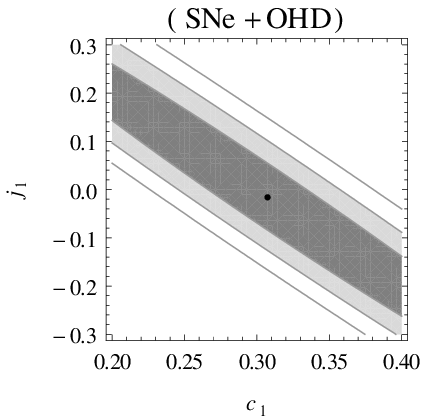}
\includegraphics[angle=0, width=0.2\textwidth]{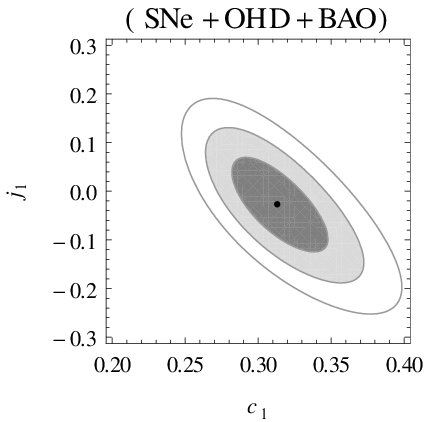}\\
\includegraphics[angle=0, width=0.2\textwidth]{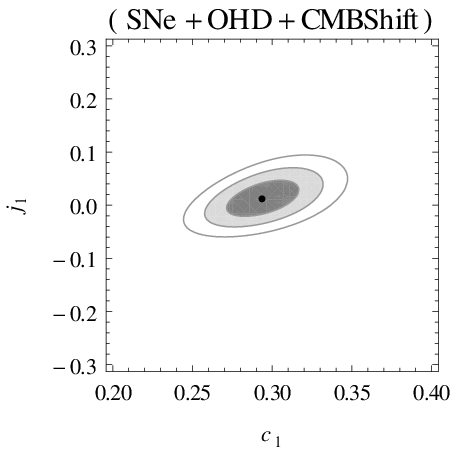}
\includegraphics[angle=0, width=0.2\textwidth]{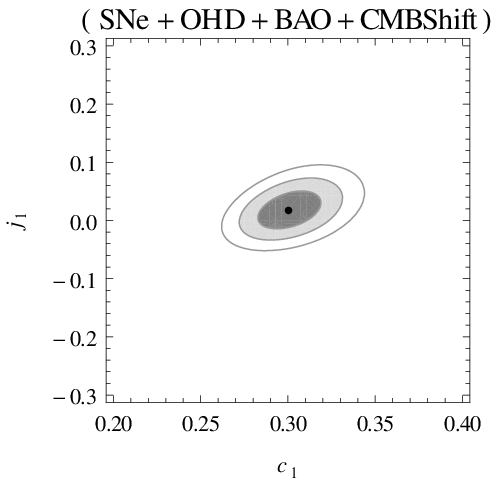}
\end{center}
\caption{{\small The confidence contours on 2D parameter space of model III. The 1$\sigma$, 2$\sigma$ and 3$\sigma$ confidence contours are presented from inner to outer regions and the central black dots represents the corresponding best fit points. The upper left panel is obtained for (SNe+OHD), upper right panel is for (SNe+OHD+BAO), lower left panel is for (SNe+OHB+CMBShift) and lower right panel is for (SNe+OHD+BAO+CMBShift).}}
\label{jz3contours}
\end{figure}

\begin{figure}[tb]
\begin{center}
\includegraphics[angle=0, width=0.2\textwidth]{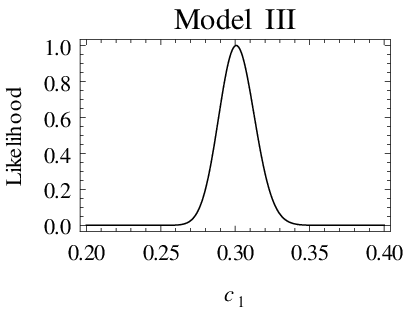}
\includegraphics[angle=0, width=0.2\textwidth]{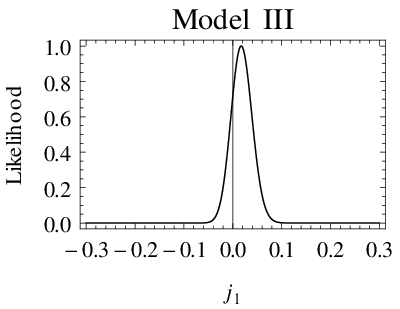}
\end{center}
\caption{{\small The marginalised likelihood functions of the model III obtained for SNe+OHD+BAO+CMBShift.}}
\label{jz3likelihood}
\end{figure}

\begin{figure}[tb]
\begin{center}
\includegraphics[angle=0, width=0.2\textwidth]{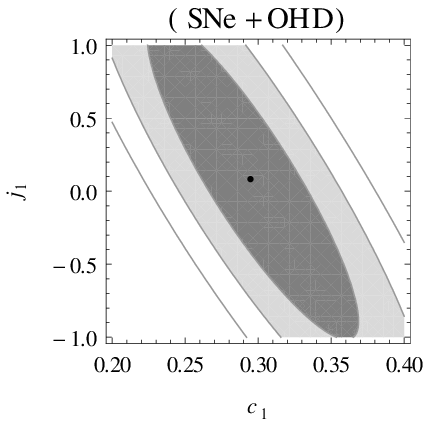}
\includegraphics[angle=0, width=0.2\textwidth]{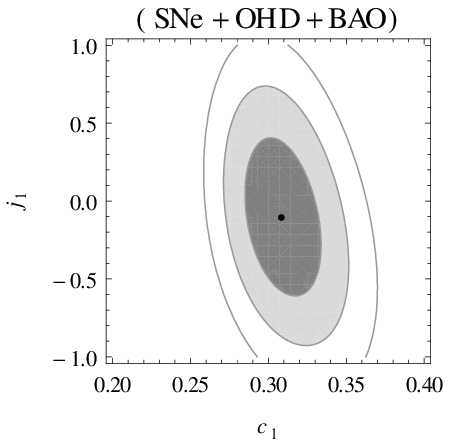}\\
\includegraphics[angle=0, width=0.2\textwidth]{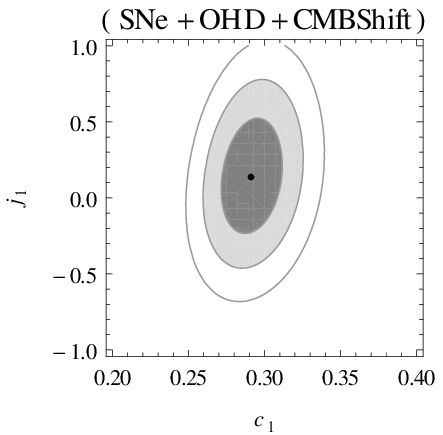}
\includegraphics[angle=0, width=0.2\textwidth]{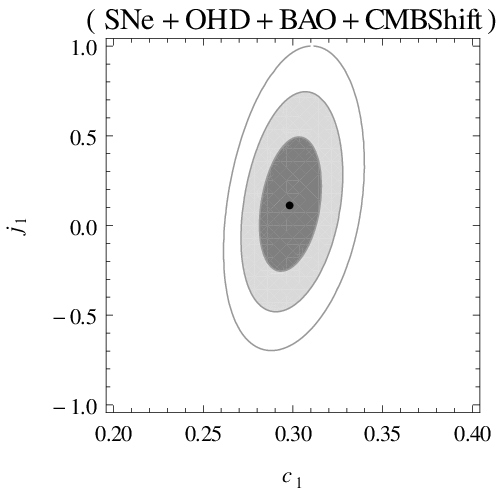}
\end{center}
\caption{{\small  The confidence contours on 2D parameter space of model IV. The 1$\sigma$, 2$\sigma$ and 3$\sigma$ confidence contours are presented from inner to outer regions and the central black dots represents the corresponding best fit points. The upper left panel is obtained for (SNe+OHD), upper right panel is for (SNe+OHD+BAO), lower left panel is for (SNe+OHB+CMBShift) and lower right panel is for (SNe+OHD+BAO+CMBShift).}}
\label{jz4contours}
\end{figure}

\begin{figure}[tb]
\begin{center}
\includegraphics[angle=0, width=0.2\textwidth]{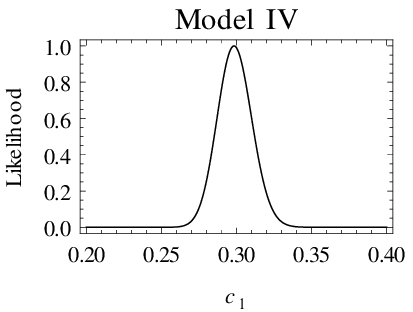}
\includegraphics[angle=0, width=0.2\textwidth]{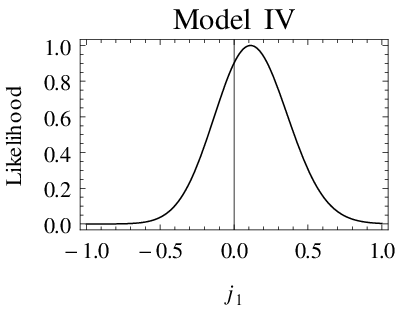}
\end{center}
\caption{{\small The marginalised likelihood functions of the model IV obtained for SNe+OHD+BAO+CMBShift.}}
\label{jz4likelihood}
\end{figure}

 Similarly the figure \ref{jz2contours} presents the confidence contours on the parameter space for model II and figure \ref{jz2likelihood} shows the marginalised likelihood functions. In table \ref{tablejzM2} the results of the statistical analysis are presented. Figure \ref{jz3contours} and \ref{jz3likelihood} are of model III and table \ref{tablejzM3} presents the results of corresponding statistical analysis. And figure \ref{jz4contours} and \ref{jz4likelihood} and table \ref{tablejzM4} correspond to model IV.
\par All the models show that the addition of CMB shift parameter data leads to tighter constraints on the model parameters. All the likelihood function plots are well fitted to Gaussian distribution. As the model parameter $j_1$ indicates the deviation of the models from $\Lambda$CDM (for $\Lambda$CDM $j_1=0$), it is imperative to note that $\Lambda$CDM remains within the $1\sigma$ confidence regions of all the models.


\begin{table*}[htb]
\caption{{\small Results of statistical analysis of  Model I}}
\begin{center}
\resizebox{0.65\textwidth}{!}{  
\begin{tabular}{ c ||c |c c c } 
 \hline
 Data  & $\chi^2_{min}/d.o.f.$ & $c_1 $ & $j_1$ \\ 
 \hline
 \hline
 SNe+OHD & $47.02/53$ & $0.295\pm0.052$ & $0.045\pm0.628$\\ 
 \hline
  SNe+OHD+BAO & $47.11/53$ & $0.309\pm0.012$ &  $-0.080\pm0.222$\\ 
 \hline
   SNe+OHD+CMBShift & $47.03/51$ & $0.292\pm 0.012$ &  $0.080\pm0.145$\\ 
 \hline
   SNe+OHD+BAO+CMBShift & $47.99/51$ & $0.298\pm0.010$ &  $0.078\pm0.140$\\ 
 \hline
\end{tabular}
}
\end{center}

\label{tablejzM1}
\end{table*}


\begin{table*}[htb]
\caption{{\small Results of statistical analysis for Model II}}
\begin{center}
\resizebox{0.65\textwidth}{!}{  
\begin{tabular}{ c ||c |c c c } 
 \hline
 Data  & $\chi^2_{min}/d.o.f.$ & $c_1 $ & $j_1$ \\ 
 \hline
 \hline
 SNe+OHD & $47.03/53$ & $0.299\pm0.050$ & $0.002\pm0.267$\\ 
 \hline
  SNe+OHD+BAO & $47.08/53$ & $0.310\pm0.012$ &  $-0.051\pm0.093$\\ 
 \hline
   SNe+OHD+CMBShift & $47.04/51$ & $0.292\pm 0.009$ &  $0.038\pm0.051$\\ 
 \hline
   SNe+OHD+BAO+CMBShift & $47.85/51$ & $0.299\pm0.008$ &  $0.045\pm0.050$\\ 
 \hline
\end{tabular}
}
\end{center}

\label{tablejzM2}
\end{table*}

\begin{table*}[htb]
\caption{{\small Results of statistical analysis for different Model III}}
\begin{center}
\resizebox{0.65\textwidth}{!}{  
\begin{tabular}{ c ||c |c c c } 
 \hline
 Data  & $\chi^2_{min}/d.o.f.$ & $c_1 $ & $j_1$ \\ 
 \hline
 \hline
 SNe+OHD & $47.02/53$ & $0.307\pm0.100$ & $-0.016\pm0.205$\\ 
 \hline
  SNe+OHD+BAO & $47.06/53$ & $0.313\pm0.015$ &  $-0.027\pm0.045$\\ 
 \hline
   SNe+OHD+CMBSfiht & $47.04/51$ & $0.294\pm 0.011$ &  $0.012\pm0.016$\\ 
 \hline
   SNe+OHD+BAO+CMBShift & $47.60/51$ & $0.300\pm0.008$ &  $0.017\pm0.015$\\ 
 \hline
\end{tabular}
}
\end{center}

\label{tablejzM3}
\end{table*}
\begin{table*}[htb]
\caption{{\small Results of statistical analysis for  Model IV}}
\begin{center}
\resizebox{0.65\textwidth}{!}{  
\begin{tabular}{ c ||c |c c c } 
 \hline
 Data  & $\chi^2_{min}/d.o.f.$ & $c_1 $ & $j_1$ \\ 
 \hline
 \hline
 SNe+OHD & $47.02/53$ & $0.295\pm0.034$ & $0.083\pm0.510$\\ 
 \hline
  SNe+OHD+BAO & $47.13/53$ & $0.308\pm0.011$ &  $-0.106\pm0.237$\\ 
 \hline
   SNe+OHD+CMBShift & $47.02/51$ & $0.291\pm0.009$ &  $0.137\pm0.176$\\ 
 \hline
   SNe+OHD+BAO+CMBShift & $48.06/51$ & $0.298\pm0.008$ &  $0.112\pm0.176$\\ 
 \hline
\end{tabular}
}
\end{center}

\label{tablejzM4}
\end{table*}

\begin{figure}[tb]
\begin{center}
\includegraphics[angle=0, width=0.2\textwidth]{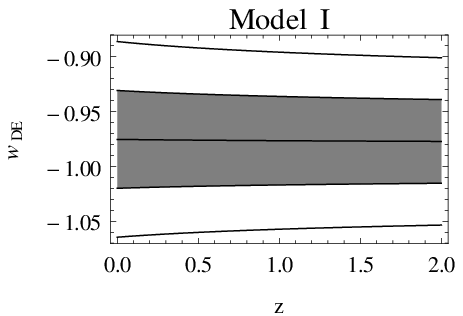}
\includegraphics[angle=0, width=0.2\textwidth]{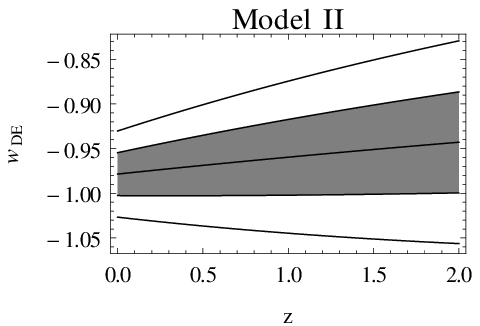}
\includegraphics[angle=0, width=0.2\textwidth]{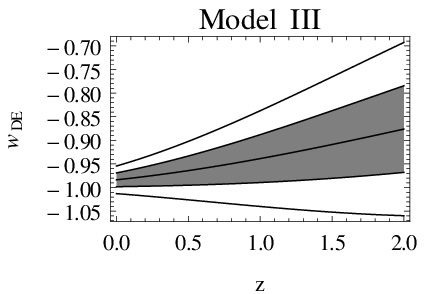}
\includegraphics[angle=0, width=0.2\textwidth]{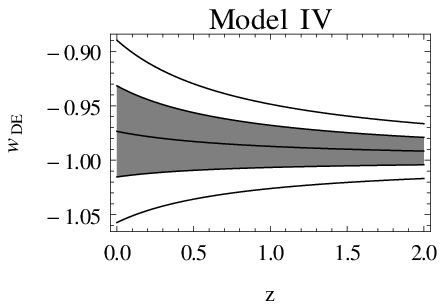}
\end{center}
\caption{{\small Plots of dark energy equation of state parameter $w_{DE}(z)$ against redshift $z$ for different parametrizations of jerk parameter $j(z)$. The 1$\sigma$ and 2$\sigma$ confidence regions, obtained from combined $\chi^2$-analysis, have been shown and the central  dark line represents the best fit curve.}}
\label{jzwzplot}
\end{figure}

\begin{figure}[tb]
\begin{center}
\includegraphics[angle=0, width=0.2\textwidth]{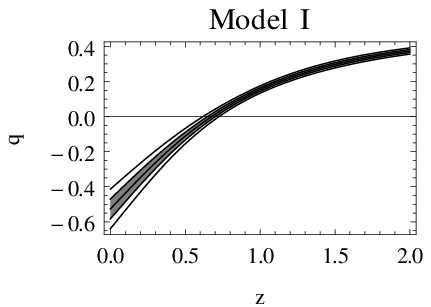}
\includegraphics[angle=0, width=0.2\textwidth]{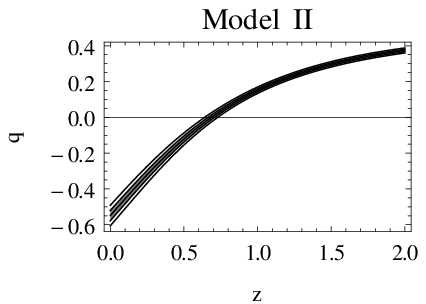}
\includegraphics[angle=0, width=0.2\textwidth]{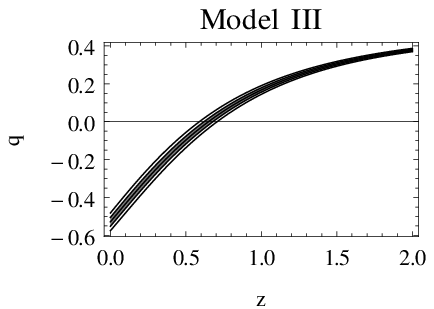}
\includegraphics[angle=0, width=0.2\textwidth]{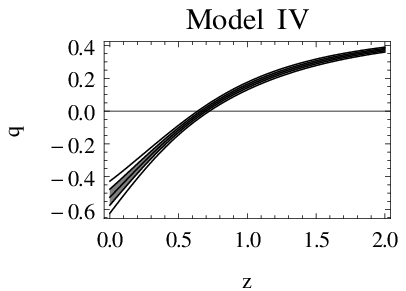}
\end{center}
\caption{{\small Plots of deceleration parameter $q(z)$ against redshift $z$ for different parametrizations of jerk parameter $j(z)$. The 1$\sigma$ and 2$\sigma$ confidence regions, obtained from combined $\chi^2$-analysis, have been shown and the central  dark line represents the best fit curve.}}
\label{jzqzplot}
\end{figure}

\begin{figure}[tb]
\begin{center}
\includegraphics[angle=0, width=0.2\textwidth]{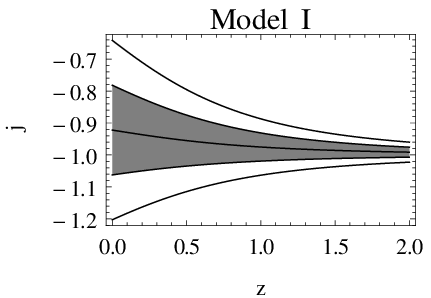}
\includegraphics[angle=0, width=0.2\textwidth]{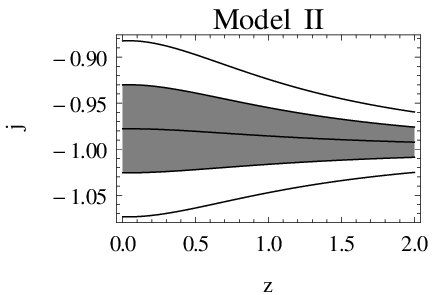}
\includegraphics[angle=0, width=0.2\textwidth]{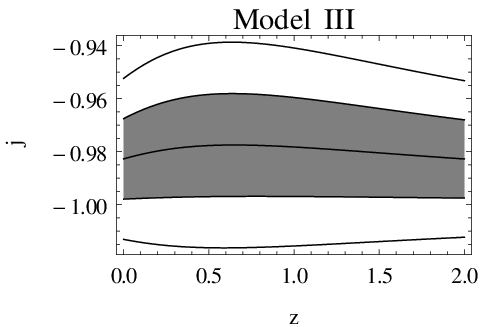}
\includegraphics[angle=0, width=0.2\textwidth]{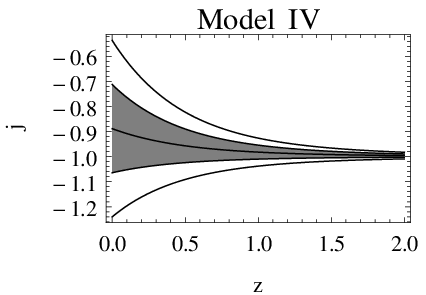}
\end{center}
\caption{{\small Plots of jerk parameter $j(z)$ against redshift $z$ for different parametrizations of jerk parameter $j(z)$. The 1$\sigma$ and 2$\sigma$ confidence regions, obtained from combined $\chi^2$-analysis, have been shown and the central  dark line represents the best fit curve.}}
\label{jzjzplot}
\end{figure}
\begin{figure*}[htb]
\begin{center}
\includegraphics[angle=0, width=0.2\textwidth]{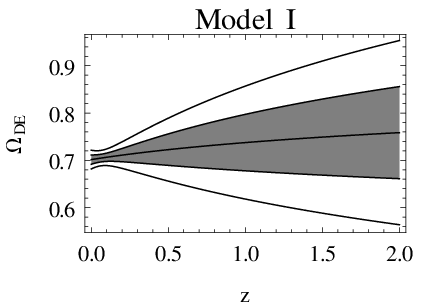}
\includegraphics[angle=0, width=0.2\textwidth]{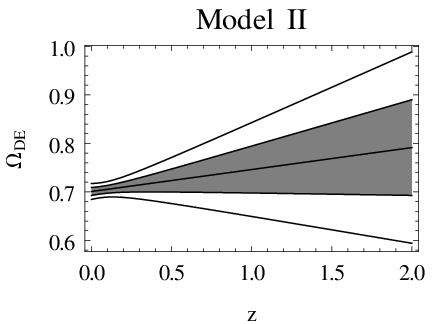}
\includegraphics[angle=0, width=0.2\textwidth]{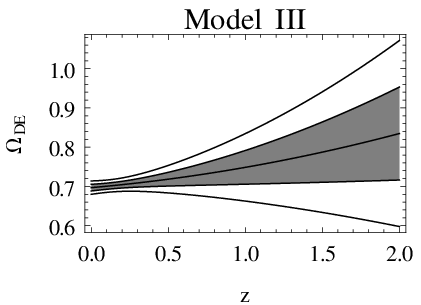}
\includegraphics[angle=0, width=0.2\textwidth]{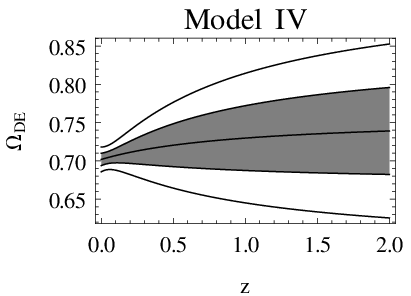}\\
\includegraphics[angle=0, width=0.2\textwidth]{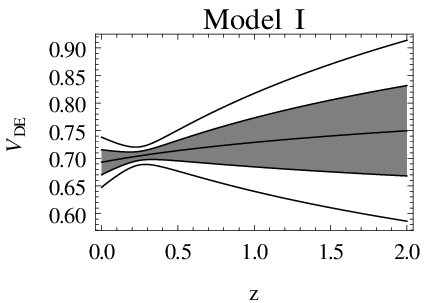}
\includegraphics[angle=0, width=0.2\textwidth]{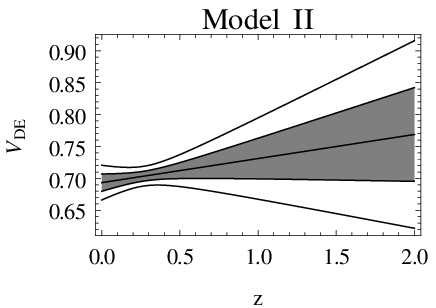}
\includegraphics[angle=0, width=0.2\textwidth]{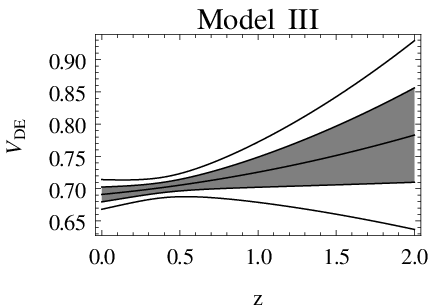}
\includegraphics[angle=0, width=0.2\textwidth]{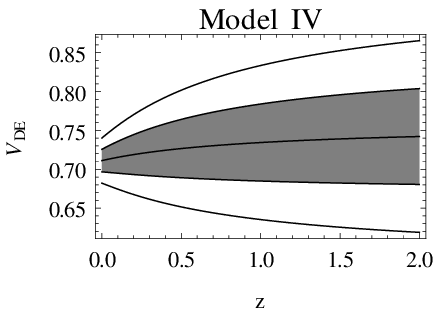}
\end{center}
\caption{{\small  Plots of dark energy density parameter ($\Omega_{DE}$) (upper panels) and the quintessence potential (lower panels) as functions of redshift $z$ for the models discussed in the present work.}}
\label{OmegaDEVz}
\end{figure*}

\vskip 20 cm

Figure \ref{jzwzplot} shows the plots of dark energy equation of state parameter as a function of redshift $z$ and figure \ref{jzqzplot} presents the plots of deceleration parameter $q(z)$ for different models discussed in the present work. The deceleration parameter plots clearly show that the models successfully generate the late time acceleration along with the decelerated expansion in the past. The plots show the transition from decelerated to accelerated expansion phase took place in the redshift range $0.6<z<0.8$. This is consistent with the recent analysis by Farooq and Ratra \cite{farooqratra} where constraints on the transition redshift are achieved for different dark energy scenario using observational Hubble data. All the models presented in this work show the dark energy equation of state parameter to  be almost constant and the best fit values at present are slightly higher than $-1$, meaning the models prefer the non-phantom nature of dark energy. 
\par Figure \ref{jzjzplot} shows the plots of jerk parameter $j(z)$ for the models. It is interesting to note that $j(z)$ is allowed by the models to take values different from that in the case of $\Lambda$CDM to start with. All the four different models show a tendency close the $\Lambda$CDM  along with a range of possibilities for $j(z)$ at the present epoch, in the 2$\sigma$ confidence region.

\section{A Bayesian analysis}
The analysis presented shows that there is hardly any preference regarding the selection of the best model from the four discussed in this work.
\par
The two commonly used criteria for model selection are  Akaike Information Criterion (AIC) \cite{akaike} and  Bayesian Information Criterion (BIC)\cite{schwarz}. They are defined as
\be
AIC=-2\log{{\mathcal L}_{max}}+2\kappa,
\label{aic}
\ee
\be
BIC=-2\log{{\mathcal L}_{max}}+2\kappa\log{N},
\label{bic}
\ee
where ${\mathcal L}_{max}$ is the maximum likelihood, $\kappa$ is the number of free parameters in  the model and $N$ is the number of data points used for the statistical analysis of the model. We note that these two criteria  can hardly provide any information regarding the model selection amongst the four presented here because the values of $\chi^2_{min}$ do not differ significantly for different models and all the models have same number of parameters as well as same number of data points have been used for the statistical analysis of the models. 
\par
As the there is hardly any model to choose based on the information criteria, it is thus useful to look for an Evidence estimate for the model selection. The Bayesian Evidence $E$ is defined as 
\begin{equation}
 \label{bayesian}
E = \int {(Likelihood \times Prior) d{\theta}_{1} d{\theta}_{2}},
\end{equation}
where ${\theta}_{1}$ and ${\theta}_{2}$ are model parameters. With a flat prior approximation, the evidences calculated for the models presented are \\

Model I:   $ E_{1} = P \times 1.81358 \times 10^{-13}$, \\

Model II:  $ E_{2} = P \times 8.79425 \times 10^{-14}$, \\

Model III: $ E_{3} = P \times 2.65140 \times 10^{-14}$, \\

Model IV:  $ E_{4} = P \times 3.05074 \times 10^{-13}$, \\

where $P$ is the constant prior. This evidences show that there is hardly any model, amongst the four presented, does better than any of the other three. However, if there is any one to choose amongst these, model IV is the one which does marginally better than the other three.

\section{Discussion}
The present work deals with a parametric reconstruction of the jerk parameter $j$ which is the dimensionless representation of the third order time derivative of the scale factor. As the deceleration parameter $q$ is now an observational parameter and found to be evolving, jerk, amongst the kinematical quantities, appears to be the natural choice as the parameter of interest as this determines the evolution of $q$. The philosophy is to build up the model from the evolution history of the universe. As such this is just another way of reconstruction, but it might indicate about the nature of matter distribution and the possible interaction amongst them without any assumption on them {\it a priori}. This may paticularly be useful in the absence of a clear verdict in favour of any model. \\

The formalism proposed  by  Zhai {\it et al} \cite{zz} has been utilized in the present work, with a major difference that $j$ is allowed to pick up any value depending on a parameter to be fixed by the data as opposed to the work of Zhai {\it et al} where $j$ is constrained to mimic a $\Lambda$CDM at the present epoch given by $z=0$. One interesting feature of this formalism is that the matter density parameter ($\Omega_{m0}$) automatically selects itself as a model parameter.

\par The plots of the dark energy equation of state parameter ($w_{DE}$) and the deceleration parameter ($q$) for the proposed models (figure \ref{jzwzplot} and figure \ref{jzqzplot} respectively) clearly show that the models can successfully generate  late time acceleration along with a decelerated expansion in the past. The range of redshift of transition from decelerated to accelerated expansion as indicated in the present work is consistent with the result of a recent analysis by Farooq and Ratra \cite{farooqratra}. The model parameter $j_1$ is an indicator  of the deviation of the model from $\Lambda$CDM. For all the four models, the best fit present value of jerk parameter estimated from the observational data are slightly greater than $-1$ and has $j=-1$ within 1$\sigma$ confidence region. Thus all these models are very close to the $\Lambda$CDM, but with an inclination towards a non-phantom nature of dark energy.

\par The values estimated for the parameter $c_1$, which is equivalent to the matter density parameter, are consistent with the results of the recent analysis of $\Lambda$CDM and $w$CDM models using the CMB temperature anisotropy and polarization data along with the other non-CMB data\cite{xiali}.

\par  A constant value of jerk is in fact allowed in all the four models within 1$\sigma$ confidence level (figure \ref{jzjzplot}). But the particular value estimated by Rapetti {\it et al}\cite{rapetti} is out of 1$\sigma$ confidence region of the present models. An evolving jerk parameter  had been discussed by Zhai {\it et al} \cite{zz} where only the supernova distance modulus data (SNe) and observational Hubble data (OHD) were used for the statistical analysis of the models.  In the present work, though the same mathematical formulation has been used as Zhai {\it et al}, tighter constraints on the parameter $j_1$ have been achieved by introducing the BAO and CMB shift parameter data along with SNe and OHD.

\par We can look for a quintessence potential from the present analysis. Let us assume that in equation (\ref{friedmann1}) and (\ref{friedmann2}), $\rho_{DE}$ and $p_{DE}$ are assumed to be given by a quintessence scalar field as $\rho_{DE}=\frac{1}{2}\dot{\phi}^2+V(\phi)$ and $p_{DE}=\frac{1}{2}\dot{\phi}^2-V(\phi)$ where $\phi$ is the quintessence field and $V=V(\phi)$ is the associated potential. As the analytic expressions of $\rho_{DE}$ and $p_{DE}$ for the models can be obtained with the aid of equation (\ref{rhoDE}) and (\ref{pDE}), the evolution of the quintessence potential $V=V(z)$ can be figured out for the reconstructed models. The upper panels of figure \ref{OmegaDEVz} show the variations of $\Omega_{DE}$, the dimensionless dark energy density parameter. The present value is close to 0.7 for all the four models. The lower panels of figure \ref{OmegaDEVz} show the evolution of the potential V, scaled by critical density ($3H_0^2/8\pi G$), as a function of $z$. The
best fit of the potential remains almost constant, in the range $0<z<2$, the upper limit of $z$ being chosen substantially  above the redshift of transition from decelerated to accelerated phase of expansion ($q=0$ between $z=0.6$ and $z=0.8$). So at least in this range of $z$, the potential is neither freezing nor thawing \cite{callinder,saherresen} but rather a constant, leading to a  slow-roll scalar field.

\par The systematic uncertainties of supernova observations are considered in the statistical analysis presented here as  some of them might have their say as well on the results, such as the colour-luminosity parameter might depend on the redshift, and hence affects the magnitude in the analysis of Supernova data \cite{wangsy}. There are some recent discussion on the effects of systematics which may be worthwhile in any analysis\cite{rubin, shaferhuterer}. But we have made an attempt to use data sets which are either uncorrelated or the correlation is rather low. It deserves mention that CMB data has been used to remove the dependence of the sound horizon in the case of the BAO data. The measurement of the acoustic scale $l_{A}$ and the CMB shift parameter $R$ are somewhat correlated. This correlation, calculated from the normalized covariance matrix given by Wang and Wang\cite{wangwang}, is not too large and not likely to change the results significantly. So this correlation is ignored in the present work.

\par Very much like the previous exhaustive work on the reconstruction of the jerk paramater by Zhai {\it et al}\cite{zz}, the present reconstruction also incorporates the $\Lambda$CDM model well within the $1\sigma$ error bar. But the major difference is that Zhai {\it et al} forced the jerk parameter to mimic the $\Lambda$CDM through their parametrization at the present epoch ($z = 0$), but the present work relaxes that requirement and finds that model is inclined away, though not in a big way, towards a non-phantom behaviour.

\par The main conclusion, therefore, is that the $\Lambda$CDM is very close to be the winner as the candidate for the favoured model with a marginal inclination towards a non-phantom behaviour of the universe. However, the present work deals with situations each of which yields the $\Lambda$CDM model as a special case ($j_{1} = 0$). Anyway, a high departure from the $\Lambda$CDM has not been ruled out ab inito, the reconstructed value of $j_{1}$ shoulders the task of the determination of the departure. A drastically different form of $j$ may be attempted in a future work.


\end{document}